\pdfoutput=1
\documentclass[prb,twocolumn,amsmath,amssymb,amsfonts,letterpaper,superscriptaddress]{revtex4}
\newcommand{\be}{\begin{equation}}
\newcommand{\ee}{\end{equation}}

\usepackage{graphicx,amsmath}
\usepackage{dcolumn}

\newcommand{\ltsim}{\protect\raisebox{-0.5ex}{$\:\stackrel{\textstyle <}
        {\sim}\:$}}
\def\go{G\=o}

\newcommand{\fig}[4]
{
\begin{figure}[#4]
\resizebox{\columnwidth}{!}{\includegraphics{#1}}
\caption{\label{#3}#2}          
\end{figure}
}

\newcommand{\figstar}[5]
{
\begin{figure*}[#4]
\resizebox{#5 \columnwidth}{!}{\includegraphics{#1}}
\caption{\label{#3}#2}          
\end{figure*}
}

\newcommand{\figfig}[7]
{
\begin{figure}[#7]
\resizebox{#2\columnwidth}{!}{\includegraphics{#1}}
\resizebox{#4\columnwidth}{!}{\includegraphics{#3}}
\caption{\label{#6}#5}          
\end{figure}
}

\begin{document}

\title{The limited role of non-native contacts in folding pathways of a lattice protein}

\author{Brian C. Gin}

\affiliation{Department of Chemistry, University of California at Berkeley, 
Berkeley, California 94720}
\affiliation{Chemical Sciences and Physical Biosciences Divisions, Lawrence Berkeley National
Lab, Berkeley, California 94720}
\affiliation{School of Medicine, University of California at San Francisco,
San Francisco, California 94143} 
\author{Juan P. Garrahan}

\affiliation{School of Physics and Astronomy, University of Nottingham, Nottingham, NG7 2RD, U.K.}

\author{Phillip L. Geissler\footnote{Corresponding author. E-mail: geissler@berkeley.edu}}
\affiliation{Department of Chemistry, University of California at Berkeley, 
Berkeley, California 94720}
\affiliation{Chemical Sciences and Physical Biosciences Divisions, Lawrence Berkeley National
Lab, Berkeley, California 94720} 
\begin{abstract}
Models of protein energetics which neglect interactions between amino acids that are not adjacent in the native state, such as the \go~model, encode or underlie many influential ideas on protein folding. Implicit in this simplification is a crucial assumption that has never been critically evaluated in a broad context: Detailed mechanisms of protein folding are not biased by non-native contacts, typically imagined as a consequence of sequence design and/or topology. Here we present, using computer simulations of a well-studied lattice heteropolymer model, the first systematic test of this oft-assumed correspondence over the statistically significant range of hundreds of thousands of amino acid sequences, and a concomitantly diverse set of folding pathways. Enabled by a novel means of fingerprinting folding trajectories, our study reveals a profound insensitivity of the order in which native contacts accumulate to the omission of non-native interactions. Contrary to conventional thinking, this robustness does not arise from topological restrictions and does not depend on folding rate. We find instead that the crucial factor in discriminating among topological pathways is the heterogeneity of native contact energies. Our results challenge conventional thinking on the relationship between sequence design and free energy landscapes for protein folding, and help justify the widespread use of \go-like models to scrutinize detailed folding mechanisms of real proteins.
\end{abstract}

\maketitle

\section{Introduction}
Current understanding of protein folding has been strongly shaped by theoretical and
computational studies of simplified models \citep{Shakhnovich2006}. Such models are
typically constructed by discarding fine details of molecular structure or by making
simplifying assumptions about the energies of interaction among amino acid residues. A
special class of models, based on \go's insights \citep{Go1983}, asserts that only a
subset of interactions, those between segments of a protein that contact one another
in the native state, are crucially important for folding. The G\=o model further
assumes a unique energy scale for these native contacts. Here, we will focus on
elaborated ``\go-like'' models that allow for a diversity of native contact energies.

Neglect of non-native contacts offers substantial computational relief to numerical
simulations, allowing thorough kinetic and thermodynamic studies to be performed even
for detailed molecular representations
\cite{Shimada2000,Shimada2001,Takada1999,Shoemaker1999}. It further establishes a
basis for theories that focus on gaps in the spectrum of conformational energies
\cite{Shakhnovich1990,Sali1994a} and the funnel-like nature of potential energy
landscapes \cite{Bryngelson1987,Onuchic1995,Onuchic1996,Onuchic1997,Socci1998}.
Corroborated by experiment, concepts intrinsic to and inspired by \go-like models now
form a canon of widely accepted ideas about how proteins fold
\cite{Shakhnovich2006,Pande1998,Onuchic2004}.

The G\=o model was originally proposed as a schematic but microscopic perspective on
the stability and kinetic accessibility of proteins' native states. It accordingly
provided generic insight into issues of cooperativity, nucleation, and the
relationship between sequence and structure \cite{Shakhnovich2006}. Recent studies
have ascribed a much more literal significance to the detailed dynamical pathways
defined by \go-like models \cite{Takada1999}. In particular, direct comparisons have
been drawn between folding mechanisms predicted by \go-like models for specific
proteins and those suggested by experimental results \cite{Karanicolas2003,
Levy2006,Simler2006}. However, it is not clear to what extent such a detailed correspondence with
\go-like models should be expected. General theories offer only rough guidance, and
few computational studies have compared folding pathways of \go-like models and their
``full'' counterparts (in which non-native contact energies are included) in a broad
context \cite{Clementi2004}.




Very favorable interactions between segments of a protein that are not
adjacent in the folded state generally impede folding. They might do
so by introducing detours or traps on the route to the native state,
or simply by stabilizing the ensemble of unfolded conformations
\cite{Sali1994,Paci2002,Paci2002a}. It is often imagined that the
former possibility plagues a vast majority of non-natural amino acid
sequences, which fold sluggishly if at all \cite{Shakhnovich1993,
Gutin1998}. According to this picture, non-native contacts should
feature prominently in the convoluted folding pathways of an
undesigned sequence. Such kinetic frustration could pose several
biological risks {\em in vivo}, where aggregation and slow response
can be serious liabilities. Indeed, typical
proteins taken from living organisms fold reliably and with relative
efficiency \cite{Mirny1998}.

These notions and observations motivate a ``principle of minimum frustration''
asserting that natural amino acid sequences have been ``designed'' by evolution to
minimize the disruptive influence of non-native contacts on the dynamics of folding
\cite{Bryngelson1987}. One might thus apply \go-like models to these designed
sequences with confidence, since the omitted interactions are precisely the ones whose
effects have been mitigated by natural selection. By contrast, one might expect
\go-like models to poorly represent folding mechanisms of slowly folding molecules,
whose non-native interactions are presumably responsible for hampering pathways to the
native state \cite{Paci2002,Paci2002a}.

Testing these ideas of sequence design and kinetic frustration is made difficult by
several factors. Experimentally, microscopic details of folding kinetics cannot be
resolved but only inferred from indirect observables or the effects of mutations.
Furthermore, the most concrete hypotheses stemming from the principle of minimum
frustration involve \go-like models, which cannot be realized in the laboratory.
Computer simulations of detailed molecular representations can generate, at great
cost, dynamical information sufficient to determine a folding mechanism for only the
smallest of natural proteins \cite{Schaeffer2008}. Although the statistical dynamics
of coarse-grained or schematic representations can be readily explored, biology does
not provide collections of fast-folding and slow-folding sequences to compare in these
artificial contexts. Finally, even when appropriate ensembles of sequences and
ensembles of folding trajectories are available, useful comparison of \go-like models
and its full counterpart requires a compact way of characterizing the course of highly
chaotic dynamics \cite{Pande1999a}. A general method for this purpose is not
available, though studies of nucleation as a rate-limiting fluctuation provide a
useful starting point \cite{Abkevich1994,Sutto2006}.



This paper presents the first systematic, large-scale comparison of
folding pathways within \go-like and full models. We focus on a
schematic lattice representation of proteins, well-suited for this
task in several ways: (a) geometrically, because contacting segments
of the chain can be unambiguously identified, (b) statistically,
because representative ensembles of folding trajectories can be
generated for large numbers of amino acid sequences, and (c)
conceptually, because the essential competition between contact
energetics and chain connectivity can be isolated from complicating
effects of secondary structure, side-chain packing, etc. While these
latter effects unquestionably bear in important ways on the folding of
real proteins, it is nevertheless imperative to understand the
fundamental physical scenarios they enrich and modify. Indeed, much of
biologists' working intuition for protein folding and design was
developed in the context of similarly schematic models. Our results
challenge some of those notions.

It has been conjectured that well-designed lattice heteropolymers fold
through mechanisms that are determined solely by their native
structures \cite{Mirny1998}.  Were this hypothesis correct, for both full and \go-like
models, a comparison of fast-folding pathways in the two models would
not be especially informative. In that case the sequence of events
that advance a molecule toward the native state (which we designate as
its folding mechanism) would be exclusively a question of geometry and
local mobility. We have found, to the contrary, that a wealth of
folding mechanisms are possible even for a {\em single} native
conformation.

Spanning a range of hundreds of thousands of sequences, with widely varying rates and mechanisms,
the work reported in this paper constitutes a thorough test of certain
aspects of the principal of minimum frustration and addresses at a new
level of kinetic detail the dynamical realism that can be expected
from \go-like models. Our results for the lattice heteropolymer model
evidence a remarkably strong mechanistic correspondence between full
and \go-like models. Unexpectedly, this dynamical conformity holds not
only for fast-folding sequences but also for the slowest sequences
whose folding can be followed in practice. Close correspondence in
folding mechanisms holds as long as the \go-like approximation retains
the heterogeneity in native contact energies of the full
potential. These findings suggest a profound frustration invariance in
the ensemble of trajectories that proceed from deep within the
unfolded state all the way to the native structure.

\figstar{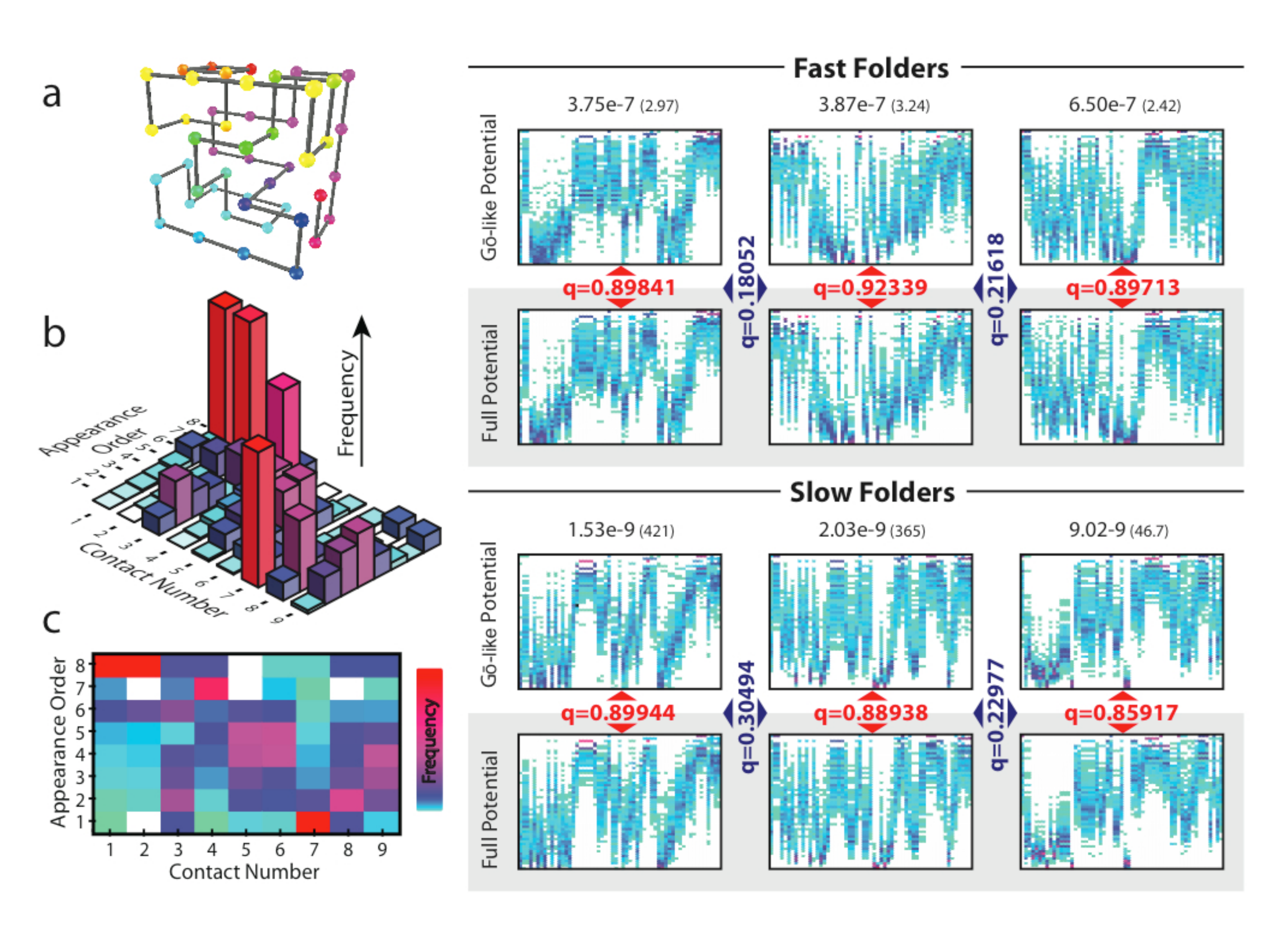}{(a) 48-mer native structure of the lattice heteropolymer studied in this work.  (b) Example of histograms of the order of permanent formation of native contacts (contact appearance order, or CAO) for each of the nine native contacts of a 12-mer lattice structure.  Histograms are collected from the set of folding trajectories of a given amino acid sequence.  (c) Same as Fig. 1b but shown as a density map.   (Right, Top)  CAO maps of three fast folding sequences of the 48-mer (Fig. 1a), for both the full potential energy and the \go-like approximation (which disregards non-native contact energies, but maintains the original heterogeneity in native contact energies).    The overlap parameter $q$ quantifies the similarity of CAO maps, and thus topological folding pathways.   The overlap of CAO maps between full and \go-like potentials for each sequence is close to one, $q \approx 0.9$, indicating the similarity of their folding mechanisms.    In contrast, the overlap between CAO maps of different sequences is much smaller, $q \ltsim 0.2$.  (Right, Bottom)  Same as before but now for three slow folding sequences.  Again, the CAO distributions of full and \go-like potentials are very similiar, while those between different sequences are not.}{fig1}{ht}{2}

\section{Methods}
We focus on lattice heteropolymers, whose folding properties have been
studied extensively for specific example sequences, structures, and
chain lengths \cite{Shakhnovich1994,Shakhnovich2006}. Here, a protein's conformation is described by a
self-avoiding walk on a three dimensional lattice with spacing
$a$ (see for example Fig 1a).  Each vertex of this walk represents an amino acid monomer, which
possesses no internal structure and interacts only with ``contacting''
monomers that occupy adjacent vertices. For a chain comprising $N$
monomers the energy of a particular configuration can thus be written
\begin{equation}
E = \sum_{i=1}^{N-1}  \sum_{j>i}^N  u_{\rm core}(r_{ij})
+ \sum_{i=1}^{N-3} \sum_{j=i+3}^N B_{ij} \Delta(r_{ij}-a) ,
\label{full}
\end{equation}
where $r_{ij}= |{\bf r}_i-{\bf r}_j|$.  The hard-core potential $u_{\rm core}(r)$, which takes on values of
$\infty$ for $r=0$ and $0$ for $r>0$, imposes the constraint of
self-avoidance. Interaction energies $B_{ij}$ are determined by the
sequence-dependent identities of monomers $i$ and $j$ according to the
model of Miyazawa and Jernigan \cite{Miyazawa1985} (MJ), and act only at a spatial separation
of one lattice spacing [$\Delta(x)=1$ if $x=0$ and vanishes
otherwise].

The standard dynamical rules for evolving such a chain molecule
proceed from a Metropolis Monte Carlo algorithm. Trial moves, in which
one or two randomly selected monomers move in an ``edge-flip'' or
``crankshaft'' fashion, are accepted with probabilities that generate
a Boltzmann distribution at temperature $T=0.16 \epsilon_{0}\,/k_{\rm
B}$, where $\epsilon_{0}$ sets the energy scale of the MJ model.  For
example, the strongest attractive interaction (between two cysteines)
has an energy $\epsilon_{\rm CC}=-1.06 \epsilon_{0}$; for
lysine-lysine $\epsilon_{\rm KK}=0.25\epsilon_{0}$. Folding
trajectories are initiated from swollen configurations drawn from a
high-temperature ($k_{\rm B}T/\epsilon_0=100$) equilibrium distribution in which contact energies are
negligible compared to typical thermal excitations.

This caricature clearly lacks many of the chemical details underlying
the function and secondary structure of real proteins. By capturing an
essential interplay between diverse local interactions and constraints
of polymer connectivity, it nonetheless recapitulates many nontrivial
features of protein statistical mechanics: Even for chains of modest
length (say, $N=27$), the number of possible conformations is
sufficiently immense to motivate Levinthal's paradox, i.e., it is not
obvious that they should be able fold at all.  Folding occurs
in a cooperative fashion, and occurs efficiently only for
well-designed sequences. For a given sequence certain residues figure
much more prominently in folding kinetics than others;
correspondingly, certain residues are more highly conserved than
others in computer simulations of evolutionary dynamics.

The \go-like approximation of the model of Eq.\ (\ref{full}) is
constructed simply by ignoring the energies of non-native
contacts, 
\begin{equation}
\tilde{E} = \sum_{i=1}^{N-1} \sum_{j>i}^N  u_{\rm core}(r_{ij})
+ \sum_{i=1}^{N-3} \sum_{j=i+3}^N {\cal N}_{ij} B_{ij} \Delta(r_{ij}-a) ,
\label{go}
\end{equation}
where ${\cal N}_{ij}=1$ if the monomers $i$ and $j$ are adjacent in the 
native configuration, and ${\cal N}_{ij}=0$ otherwise.  While disregarding the energy contribution of non-native contacts, the energy function $\tilde{E}$ of Eq.\ (\ref{go}) retains the full heterogeneity in native contacts energies of the original potential, Eq.\ (\ref{full}).  We will show below that it is a crucial aspect of the \go-like models we study here.

Many studies previously suggested that lattice heteropolymers of modest length fold
via a nucleation mechanism \cite{Abkevich1994,Sutto2006}. Formation of a handful of
key contacts poises the system at a transition state, from which the chain can rapidly
access the folded state or, with equal probability, return to the unfolded state. This
set of crucial contacts comprises a ``folding nucleus'' and serves as a bare synopsis
of dynamical pathways that lead to the native state.

A cogent comparison of folding mechanisms requires a means of
characterizing dynamical pathways that is both thorough and
computationally inexpensive. Identifying the folding nucleus satisfies
neither or these necessities well.  In particular, locating
configurations from which the folded and unfolded states are equally
accessible involves propagation of many trajectories and, by itself,
does not delineate routes toward and away from the transition state
\cite{Dokholyan2000}.  We have devised an alternative measure that is
not only succinct and computationally tractable, but also
characterizes the entire route from the unfolded to the folded state.
Specifically, we record the order in which native contacts form
permanently during a protein's folding mechanism.  Our parameters thus
chronicle lasting changes in the chain's ``topology'', understood in
terms of linkages through the polymer backbone and through non-bonded
contacts.

\figstar{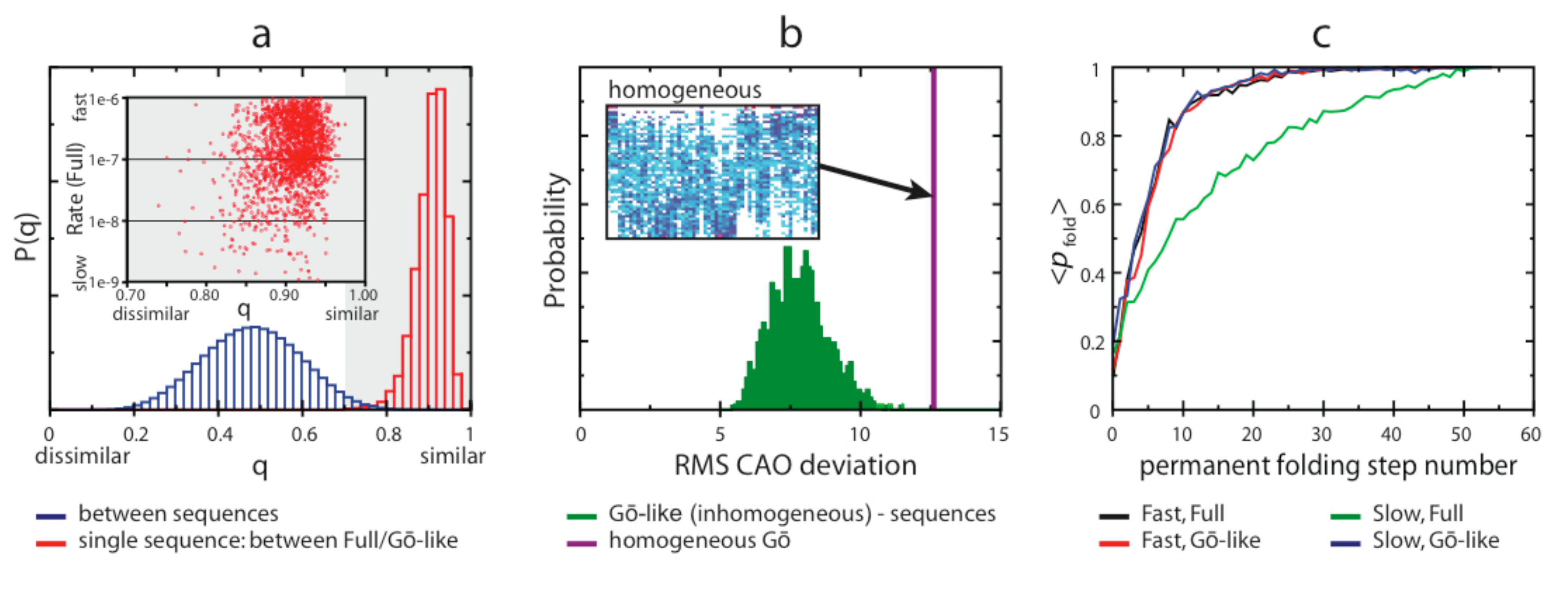}{(a) Distribution of CAO overlaps, $P(q)$, between different sequences, and between full and \go-like potential, for $1000$ sequences chosen randomly out of $10^5$ sequences that fold to the 48-mer structure of Fig. 1a. The sequences in this distribution were generated by a single high $T_{\rm ev}$ evolutionary trajectory (see Appendix).  The inset shows that the similarity between full and \go-like pathways for each sequence is independent of folding rate. Data for this inset was generated from $2000$ sequences chosen randomly from 5 independent evolutionary runs ($5\times10^5$ total sequences), all folding to the native 48-mer structure of Fig. 1a. (b) Distribution of the root-mean-squared fluctuations of contact order, $\sqrt{\delta C}$, over the set of \go-like sequences.  CAOs in heterogeneous \go-like potentials vary less from one folding trajectory to another than in the homogeneous G\=o model.  It is the heterogeneity in native contact energies that selects specific folding pathways; this selectivity is absent in a homogeneous G\=o potential.  The inset shows the CAO map of the homogeneous G\=o potential, cf. Fig. 1.
(c) Average $p_{\rm fold}$ as a function of number of permanent native contacts formed, for the full and \go-like potentials, for a fast and a slow folding sequence.  In all cases $p_{\rm fold}$ is close to zero until the first permanent contacts are made, confirming that our CAO analysis captures the relevant dynamical folding regime. $p_{\rm fold}$ is the probability for a given conformation to reach the folded state before unfolding. For a given folding trajectory, we calculate $p_{\rm fold}$ according to the method of \citet{Faisca2008}, by running independent trajectories from configurations chosen at evenly-spaced time intervals. We regard a molecule as unfolded when the instantaneous number of native contacts drops to a value consistent with the average number of native contacts in the unfolded state. Additionally, we require that this threshold lie below any value found in equilibrium fluctuations of the native state.}{fig2}{ht}{2}

This contact appearance order (CAO) is a highly nontrivial measure of
the progress toward folding and provides a detailed characterization
of mechanism in the sense we have defined. It is simple to calculate
from the time-dependence of a trajectory spanning unfolded and folded
states. Like persistence times \cite{RitortS03} 
in the context of non-equilibdium systems, such as glasses, it 
is intrinsically a multi-time quantity; it
can neither be computed for a single configuration, nor can it be used
to build constrained ensembles whose statistics shed light on the
nature of reaction coordinates. But, also like persistence times \cite{RitortS03}, it
focuses attention on key dynamical events with unmatched
precision. For our purpose of diagnosing the occurrence of lasting
topological changes, CAOs serve almost ideally. For some other
approaches, e.g., surveying the free energy landscapes on which
folding takes place, CAOs would serve poorly.

We have verified that the mechanistic meaning we ascribe to CAOs is
consistent with more conventional characterizations of reaction
progress.  Most importantly, the order of a contact's appearance
correlates strongly with a statistical measure of commitment to
folding at the time when that contact forms permanently. We use the
parameter $p_{\rm fold}$, the probability that a trajectory initiated
from a given configuration will reach the folded state before first
relaxing to a state with few native contacts \cite{Faisca2008}, to
demonstrate this fact. Fig. 3c shows
that the average value of $p_{\rm fold}$ rises steadily with CAO,
from a value well below $p_{\rm fold}=1/2$ up to $p_{\rm fold}=1$.

The point at which $p_{\rm fold}$ crosses $1/2$ is often considered
the transition state for folding. The set of contacts consistenly
present in such configurations is correspondingly designated as the
folding nucleus.  We have confirmed that the nucleus identified in
this way corresponds closely with the set of contacts that have formed
permanently when $p_{\rm fold}=1/2$. Additionally, we have verified that 
the CAO-identified nucleus of several sequences from \citet{Mirny1998} 
are consistent with the nucleus identified in that study.
While this consistency check reflects
favorably on the soundness of exploring folding mechanisms by
scrutinizing CAOs, it does not imply that CAO analysis is predicated
on putative nucleation mechanisms for folding. Regardless of whether
the rate-determining steps in folding are uphill, downhill, or neutral
in free energy; regardless of whether folding is kinetically a
two-state phenomenon; regardless of whether the progress of folding is
plagued by long-lived kinetic traps, CAOs trace a history of
conformational change that emphasizes any event with enduring
topological consequences.

What CAOs do not resolve is the unproductive development of native
structure. Attention is focused solely on segments of time evolution
that bridge folded and unfolded basins of attraction.  Occasional
excursions within the unfolded state amass an atypically large number
of native contacts, but due either to topology or to the presence of
interfering non-native contacts do not in fact make progress toward
folding. CAOs contain no information about these excursions. In
comparing full and \go-like models, we therefore make no statements
about the character of such non-folding dynamics.  By exclusively
examining accumulation of native contacts, we also lose direct
information regarding the evolution of non-native contacts.  If the
rupture of a particular non-native contact were a crucial step in
folding of a certain sequence, our methods would not detect its
occurrence explicitly. We stress, however, that substantial non-native
structure is present when the first permanent native contacts are
formed. We could therefore indirectly detect the significance of
non-native contact dynamics through influences on the pattern of
early topological changes.


Compiling the order of permanent contact formation over many folding
trajectories of a given sequence, we construct for each native contact
a statistical distribution of CAO.  Fig. 1b,c illustrate how the set
of resulting CAO histograms form a visual fingerprint of a sequence's
folding mechanism.  Because the dynamical events it chronicles span a
wide range of $p_{\rm fold}$, a CAO histogram characterizes not only
the transition state for folding, but also the dynamics of ascent to
and descent from the transition state. The correspondence between an
amino acid sequence and its CAO histogram is as subtle as (if not more
so) the connection between sequence and native conformation that
defines some of the most challenging aspects of the protein folding
problem. Most of the results we will present concern a {\em single}
native structure (that shown in Fig. 1a for $N=48$), removing a
potentially trivial agreement between full and \go-like models. Even
for this unique structure, sequences of the full model differing by
only a few point mutations can exhibit qualitatively different CAO
histograms, reflecting substantial
changes in folding pathway. The distribution of contact energies can
thus play a critical and complex role in determining folding
mechanism, over and above dictating its endpoint. Given this
nontrivial relationship it would be surprising if non-native contacts
did not generally act to shape or bias CAO statistics.

The primary goal of this paper is to compare the CAO statistics of
sequences propagated using full and \go-like models. In judging their
similarities and differences, it is essential to establish for
reference how significantly CAO histograms can vary, within either
model, for sequences that fold to a common structure. As mentioned
above, others have proposed that such variations are weak, i.e., that
topology of the folded structure prescribes a nearly unique
topological route for folding. Using methods described in the
Appendix, we have generated an unprecedentedly diverse set of
sequences that fold to the same target structure within the full
model. As shown in Fig. 1 variations in CAO statistics within this set
are much more substantial than previously thought. Any success of
\go-like models in reproducing folding pathways of the full model
cannot be attributed simply to their sharing a common native
structure.

We quantify similarity of CAO statistics (for two sequences within the
same model, or for full and \go-like models with the same sequence) using
an ``overlap" parameter $q$ \cite{FischerHertz}.  Inspired by
the theory of spin glasses, we define $q$ such that $0\leq
q \leq 1$, with larger $q$ representing greater similarity.  
The analogy with spin glasses would assign
an overlap $q^{(\alpha,\beta)}$ between the CAO distributions for two
sequences $\alpha$ and $\beta$ proportional to
\begin{equation}
\frac{1}{n_{\rm max}}\sum_{n=1}^{n_{\rm max}} \sum_{C=1}^{n_{\rm max}-1} 
P_n^{(\alpha)}(C) P_n^{(\beta)}(C),
\label{qexact}
\end{equation} 
where $P_n^{(\alpha)}(C)$ is the probability that native contact $n$
is made permanently at order $C$ in a folding trajectory of sequence
$\alpha$, and $n_{\rm max}$ is the total number of native contacts.
An accurate numerical estimate of the quantity in Eq.\ (\ref{qexact}),
however, is problematic to obtain, requiring the generation of an
inordinate number of folding trajectories.  As an alternative, we
define $q$ using a closely related quantity,
\begin{eqnarray}
q^{(\alpha,\beta)} = && \frac{1}{n_{\rm max}}\sum_{n=1}^{n_{\rm max}}
\left[ 
\sqrt{2\left(\frac{\sigma^{(\alpha)}_n \sigma^{(\beta)}_n }
{(\sigma^{(\alpha)}_n)^2+(\sigma^{(\beta)}_n)^2}\right) }
\right.
\nonumber
\\
&&
\left.
\times
\exp\left(
-\frac
{\left(\langle C \rangle_n^{(\alpha)} - \langle C \rangle_n^{(\beta)}\right)^2}
{\left(\sigma^{(\alpha)}_n\right)^2+
\left(\sigma^{(\beta)}_n\right)^2}
\right)
\right],
\label{qapprox}
\end{eqnarray}
where $\langle C \rangle_n^{(\alpha)} = \sum_{n=1}^{n_{\rm max}}
P_n^{(\alpha)}(C) \, C$ is the average~CAO of contact \#$n$ for
sequence $\alpha$ and $(\sigma^{(\alpha)}_n)^2 = \sum_{n=1}^{n_{\rm
max}} P_n^{(\alpha)}(C) (C-\langle C \rangle_n^{(\alpha)})^2$ is its
variance.  Equations (\ref{qexact}) and (\ref{qapprox}) are completely
equivalent in the case of Gaussian distributed CAOs.  Even for
non-Gaussian statistics, $q^{(\alpha,\beta)}$ remains a useful,
computationally tractable, and similarly bounded measure of how
similarly two sequences fold.

\figstar{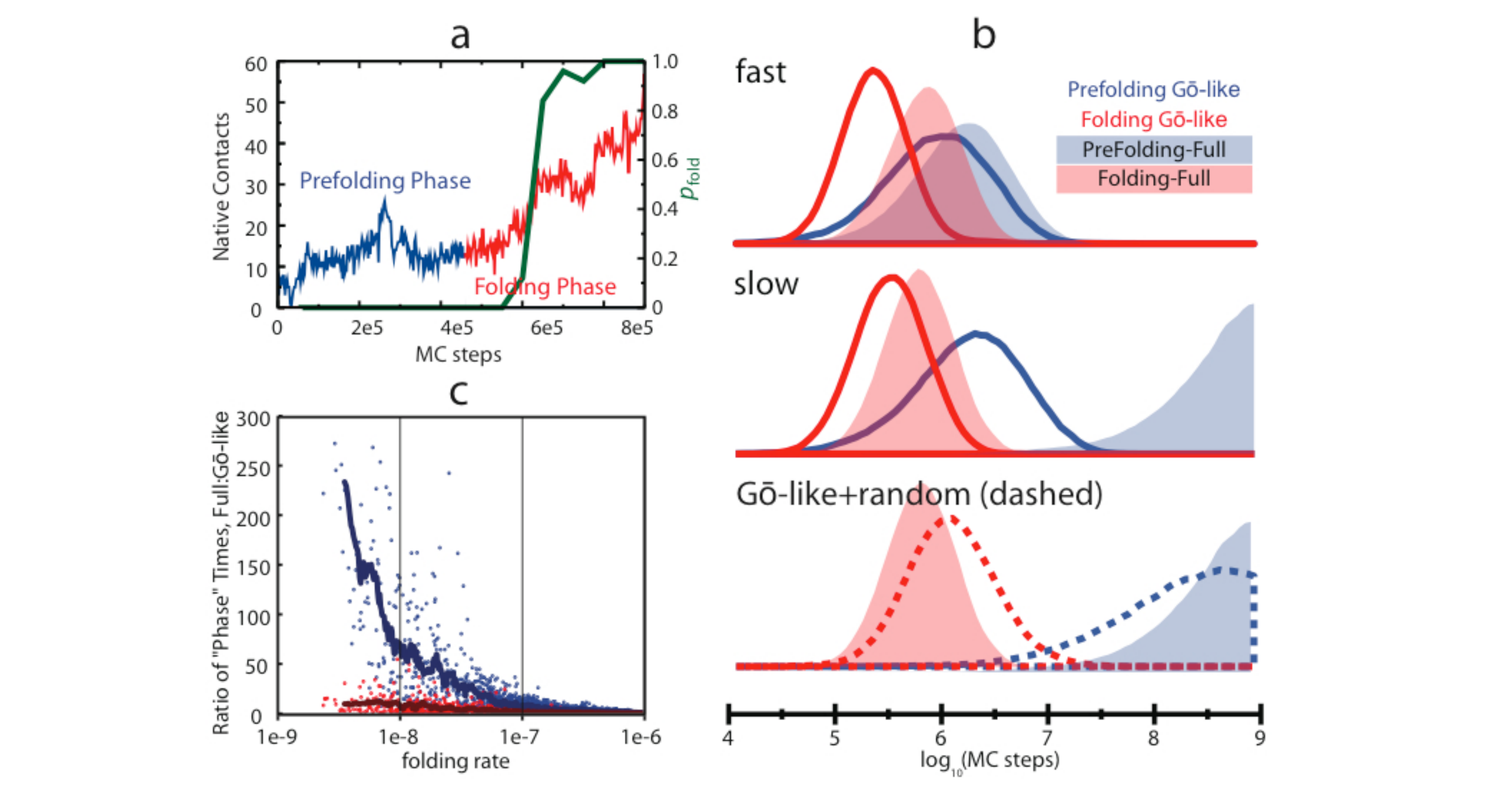}{(a) Number of native contact as a function of time in a folding trajectory, illustrating the ``prefolding''  (blue) and ``folding'' (red) phases of the dynamics.  The prefolding phase extends from the folding trajectory's start time until the time the first permanent native contact is formed.   The folding phase extends from this time to the time when the native conformation is reached.  The full (green) curve shows the $p_{\rm fold}$, which only departs from zero after the folding phase has started (cf. Fig. 2).  (b, right panels)  Distribution of the duration of the prefolding and folding phases, in the full potential and its corresponding \go-like approximation.  For fast-folding sequences (top panel) the distributions for both folding and prefolding durations of the \go-like model are close to those of the full potential.  For slow-folding sequences (middle panel) the \go-like model reproduces the distribution of folding duration, but underestimates the prefolding times.  If the \go-like potential of slow-folding sequences is supplemented by random non-native contact energies (bottom panel) the prefolding distributions can be made to mach, without disrupting the correspondence in the folding phase distributions.  (c) Ratio between full and \go-like models' folding (red) and prefolding (blue) phase durations, for all sequences ordered according to folding rate; the full lines are the average ratios for each scatter plot.  For fast folders, the average times as calculated from the full and \go-like models are comparable, both for the folding and prefolding phases.  For slow folders, the prefolding time in \go-like model is much smaller than that in the full potential, and this difference increases with decreasing folding rate.}{fig3}{ht}{2}

\section{Results and discussion}

In the ensemble of sequences we generated, the fastest folding sequences access the
native state more than 1000 times more rapidly than the slowest. CAO histograms were
generated for all sequences, each one evincing a well-defined topological pathway.
Typically, the appearance order $C$ of a given native contact varies from one
trajectory to another by only a few positions (see below). This regularity belies
substantial conformational fluctuations attending each folding event, which exert
little influence on the formation of {\em permanent} contacts. Sharply peaked CAO
histograms do not indicate a lack of complexity, but instead a successful
characterization of forward progress along the reaction coordinate for folding.

Figure 1 shows CAO histograms for several sequences folding to this specific 48-mer
structure (depicted in Fig. 1a). Results are presented for dynamics propagated
according to both full and \go-like models. Comparing these topological fingerprints
across different sequences hints at the broad variety of possible folding pathways.
Contacts essential to early stages of folding for 
one sequence can be irrelevant in the
pathway taken by another. This finding contrasts strongly with the ``one-structure
one-nucleus'' hypothesis, bolstering recent reports of dissimilar folding nuclei
\cite{Sutto2006}. 

Strong variations in the topological folding pathways chosen from one sequence to
another immediately indicate that the original homogeneous \go~model \cite{Pande1999a}
cannot capture the folding behavior of a typical sequence. With a homogeneous set of
native contact energies, that model can only discriminate between different native
structures, not between different sequences that adopt them. In loose terms folding
dynamics of the homogeneous \go~model resemble a superposition of those we determined
for diverse sequences of the full model. Whereas in the full model a typical set of
contact energies selects a well-defined folding pathway, an egalitarian set of
stabilizing energies permits broad sampling of routes to the native state.

\go-like models that embrace variety in native contact energies,
however, capture the topological pathways followed by their full model
counterparts with striking accuracy. CAO histograms obtained from full
and \go-like dynamics for any particular sequence can hardly be
distinguished, see Fig. 1. Not only are the average CAOs of each
contact nearly equivalent, but also fine details of CAO statistics are
unaffected by neglect of non-native contact energies. While previous
work hypothesized a dynamical correspondence for fast folders, the
topological conformity of full and \go-like mechanisms we observe for
slow folders is highly unexpected. 

For sequences with folding rates $\ltsim 10^{-9}$, we are unable to
harvest folding trajectories in sufficient numbers to construct CAO
histograms. According to microscopic reversibility, however,
topological routes for folding are identical to time-reversed routes
of unfolding. We have therefore extended our analysis of contact
appearance order for efficiently folding sequences to one of contact
{\em disappearance} order (CDO) for very sluggishly folding
sequences. The agreement between CDO histograms of full and \go-like
models is no less striking than that of the CAO histograms plotted in
Fig. 1, even in cases where the ``native'' state is grossly
unstable. These calculations are somewhat less straigthforward:
the order of first disappearance (CDO) is equivalent to the order of permanent appearance (CAO), but only for trajectories reaching the unfolded state {\em without} revisiting the native state. As such, they require specifying when a molecule has unfolded. For this purpose, we regard a molecule as unfolded when the instantaneous number of native contacts drops to a value consistent with the average number of native contacts in the unfolded state. Additionally, we require that this threshold lie below any value found in equilibrium fluctuations of the native state. We have verified that CAO and CDO histograms indeed match for sequences folding at moderate rates.

\figstar{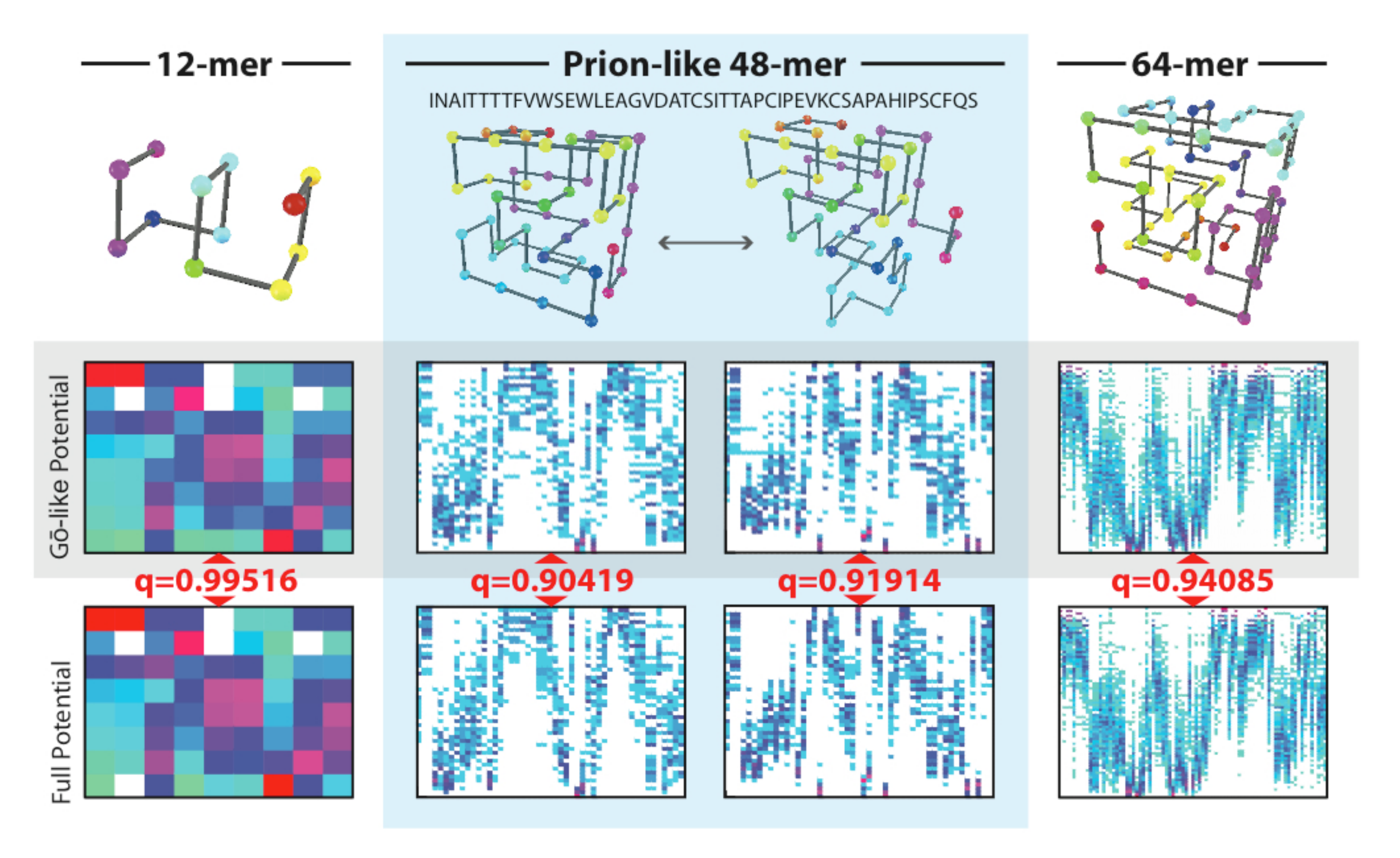}{The CAO correspondence between the full potential and the \go-like approximation is robust to changes in chain length or target native structure.  (Left) CAO maps is a 12-mer folding to the structure shown in the figure.  (Center) A sequence of the 48-mer of Fig. 1 which has a secondary stable configuration.  Each target structure defines a \go-like approximation from the set of their native contacts.  Each \go-like model predicts accurately the CAO map for folding to the corresponding structure.  (Right) Correspondence of \go-like/full CAO maps in a 64-mer. }{fig4}{ht}{2}

Quantitative measures of mechanistic diversity are presented in Fig. 2a. For each pair of
sequences generated by our evolutionary simulation we computed the similarity parameter $q$
between CAO histograms for the full model. The resulting distribution of $q$ values is
broadly peaked at $q \approx 0.4$, signifying that there is a significant diversity of CAO
pathways represented by the sequences in the ensemble. For each individual sequence we also
quantified the relationship between CAO histograms generated by full and \go-like models.
These $q$ values are distributed much more narrowly about a considerably higher average, $q
\approx 0.9$. Using sequence-to-sequence variation in CAO pathways as a yardstick, the
irrelevance of non-native contacts for the topological folding pathway is beyond doubt. The
inset to Fig. 2a emphasizes that this result has little to do with folding efficiency.
Typical $q$ values for the full/\go-like comparison are just as high for the slowest folders
examined as for the fastest.

Figure 2b quantifies the variation of CAO between folding trajectories.  For each sequence  we quantify the root mean-squared fluctuation in the contact order:
\begin{equation}
\delta C = \frac{1}{n_{\rm max}} \sum_{n=1}^{n_{\rm max}} \sqrt{\langle C_n^2 \rangle -\langle C_n \rangle^2} .
\end{equation}
Fig.\ 2b shows the distribution of $\delta C$ among the ensemble of
\go-like sequences.  It is peaked at a value of $\delta C \approx
7.5$.  In contrast, for the homogeneous \go~model $\delta C \approx
12.5$, indicating that CAO values are much more broadly distributed
between trajectories (see inset to Fig.\ 2b).  The homogenous
\go~model indeed lacks the pathway specificity exhibited when contact
energies are diverse, as in heterogeneous \go-like models.

The relevance of CAOs for the folding dynamics is illustrated in Fig.\ 2c. For two sequences
and their \go-like approximations, it plots $p_{\rm fold}$ \cite{Du1998,Dokholyan2000} as a
function of the total number of permanent native contacts formed, averaged over 200 folding
trajectories. $p_{\rm fold}$ gives the probability for trajectories initiated from a
particular configuration to fold completely before visiting the unfolded state, and provides
a standard basis for defining transition states in complex systems
\cite{Du1998,Dokholyan2000}. Fig.\ 2c shows that $p_{\rm fold} \ll 1$ when the first
permanent contact is formed. Since $p_{\rm fold}=1$ by definition when the last permanent
contact is formed, CAO histograms chronicle nearly the entire course of folding dynamics, all
the way from the unfolded basin of attraction ($p_{\rm fold}=0$) to the native state ($p_{\rm
fold}=1$).

Insensitivity of topological folding pathways to non-native contact
energies by no means implies a complete dynamical equivalence of full
and \go-like models. For example, a sequence's mean first passage time
for folding can differ by as many as three orders of magnitude for
full and \go-like models. This discrepancy is larger for sequences
with slower folding rates. Such discrepancies may be due to the
presence of off-pathway traps in the unfolded state, and possibly
non-native stabilized intermediates along the folding pathway.
However, our calculations suggest that such marked distinctions are
largely limited to dynamics occurring before the value of the
committor function $p_{\rm fold}$ increases significantly from zero,
i.e. before significant progress has been made along the folding
reaction coordinate.

As illustrated in Fig.\ 3a, we can divide each folding trajectory into
a period before any permanent contacts are made (the ``pre-folding
phase'') and the remaining period in which lasting native structure
develops (the ``folding phase''). Note that this division takes place
well before a molecule commits to the folded state ($p_{\rm fold} >
1/2$); indeed, the number of non-native contacts at the beginning of
the folding phase is typically comparable to that of the unfolded
state.  Fig.\ 3b shows the distributions of pre-folding and folding
phases' durations for two sequences representative of fast and slow
folders.  In both cases the influence of non-native contacts on the
folding phase dynamics is weak.  Non-native contacts mildly extend the
time required to complete folding after the first permanent contact is
made, by less than an order of magnitude. By contrast, pre-folding
dynamics of poorly designed sequences are quite sensitive to
non-native contact energies. For the example shown in the middle panel
of Fig.\ 3b, the waiting period prior to formation of a single
permanent contact is roughly three orders of magnitude longer in the
full model as in the \go-like model. No such dilation is observed for
sequences that fold quickly in the full model.

Because contact appearance order is a sensitive measure of approach to
the dynamical bottleneck for folding, our division of pre-folding and
folding phases is a kinetically meaningful one. Most importantly,
$p_{\rm fold} \ll 1$ throughout pre-folding dynamics as seen in
Fig. 3a, indicating that the system remains well within the unfolded
basin of attraction. Only when permanent contacts are made does
$p_{\rm fold}$ rise significantly, so that the folding phase
encompasses entirely departure from the unfolded state and transit to
the native structure. It is remarkable that non-native contacts, which
can substantially prolong dwell times in the unfolded state, exert no
discernible influence on the topological folding order, and only a
small effect on the duration of folding phase dynamics.  

Our simulations suggest that progress toward the native state
is essentially orthogonal to the formation and rupture of non-native
contacts. A number of such contacts are certainly present over much of
the course of folding, but they do little to decide what
conformational rearrangements bring a chain closer to its transition
state for folding. To further test this idea, we studied folding
dynamics governed by potential energy functions that combine aspects
of full and \go-like models.  Specifically, we selected a set of
non-native contact energies at random from a Gaussian distribution, see Fig.\ 3b.
The ``frustrating'' influence of these random energies match precisely
the behavior we have reported for the full model: CAO histograms are
completely insensitive to the average strength and variance of
non-native attractions, while overall folding rates decrease with
increasing non-native attraction strength.


The observation of correspondence between dynamics of the full lattice
model and that of a heterogeneous \go-like approximation does not noticeably
depend upon chain length or on details of native structure.  We have
generated sequences with a range of folding rates for several native
conformations of chains with lengths 8, 12, 48, and 64. For the two
shortest chains, we used each maximally compact lattice structure as a
folded state. For the two longest chains, we studied several native
structures varying significantly in compactness and in contact order
\cite{Weikl2003a}. Typical results shown in Fig.\ 4 highlight that the
fidelity of \go-like folding mechanisms is a very general feature of
these lattice heteropolymers.


\section{Conclusions}

Several arguments have been presented in the literature to justify the
use of \go~models in studying the folding mechanisms of real
proteins. Most commonly asserted (based on the principle of minimum
frustration) is that evolutionary optimization of real sequences
removes kinetic barriers and renders the energy landscape smoothly
funneled and therefore \go-like \cite{Onuchic1996,Onuchic2004}. Biases
due to topological features of the native state, unchanged in a
protein's \go-like represention, have also been invoked to justify
mechanistic fidelity \cite{Oliveira2008,Hills2008}.  Our results
demonstrate, however, that neither of these assumptions need hold for
a \go-like model to reproduce in fine detail the topological ordering
of folding events of a lattice heteropolymer.

Robustness of the detailed mechanism for folding to omission of
non-native contacts is not a consequence of sequence design within the
schematic lattice models we have studied. It is a fundamental emergent
feature of their statistical dynamics, independent of folding
efficiency over the entire range accessible to our numerical
simulations. Rather than introducing kinetic roadblocks that reshape
transition states for folding, energetic diversions due to non-native
contacts appear to strongly affect only physical properties of the
unfolded state. Even the duration of trajectory segments that span
folded and unfolded states is essentially determined by native
energies alone, despite the fact that substantial non-native structure
must be disrupted en route.

Lattice heteropolymers are perhaps the crudest representation of
protein mechanics to which our analysis could be meaningfully
applied. The correspondence between full and \go-like folding
mechanisms we have revealed might break down in more detailed
models. For example, it has been reported that lattice heteropolymers
do not exhibit glassy folding dynamics even at very low temperatures,
while non-Arrhenius temperature dependence naturally arises in
slightly elaborated models that describe side chain packing in
addition to backbone conformation \cite{Gutin1998}. \go-like
energetics could alter folding pathways by abating the frustration
underlying such glassy relaxation.  This possibility, which merits
further investigation, does not however negate the significance of our
findings. Our primary purpose is not to justify the use of \go-like
models for detailed study of real proteins' folding mechanisms. It is
instead to establish the influence of non-native interactions on
dynamics intrinsic to the fundamental interplay between chain
connectivity and heterogeneous contact interactions. That interplay,
whose understanding is central to any instructive physical picture of
protein folding, is not just present in simple lattice models -- it is
the exclusive source of their complexity. The results we have
presented therefore establish an important point: Mechanistic aspects
of protein folding that arise from the basic physics of heteropolymer
freezing are remarkably insensitive to non-native structure.

\acknowledgments
We wish to thank D. Chandler, J. Chodera, K. DuBay,
R. Jack, and S. Whitelam for useful discussions, and W. Eaton,
E. Shakhnovich, and A. Szabo for critical readings of the manuscript.

This research used resources of the National Energy
Research Scientific Computing Center, which is supported by the Office
of Science of the U.S. Department of Energy under Contract
No. DE-AC02-05CH11231. 
This work was supported by the Director, Office of Science, 
Office of Basic Energy Sciences, Chemical Sciences and Physical Biosciences 
Divisions, of the U.S. Department of Energy under Contract No. DE-AC02-05CH11231. 
In carrying out this work JPG was supported by EPSRC grant
GR/S54074/01.

\section{Appendix}

Our method of sequence generation, which effects a biased random walk in the space of all possible
sequences, is an extension of the method of \citet{Mirny1998}. To generate ensembles of sequences
folding to a specific native structure, we introduce random point mutations and accept them with a
Metropolis probability \begin{equation} P_{\rm acc} = \min\left[1,\exp\left( -{\Delta F^{\ddagger
(\beta)}-\Delta F^{\ddagger (\alpha)} \over T_{\rm ev} } \right)\right] \end{equation} that
generates a Boltzmann-like distribution. Here, $\Delta F^{\ddagger (\alpha)}$ is an estimated
activation free energy for folding of sequence $\alpha$, $k^{(\alpha)} = k_0 \exp(-\Delta
F^{\ddagger (\alpha)}/k_{\rm B}T) $. We estimate the folding rate constant $k^{(\alpha)}$ for
sequence $\alpha$, relative to the rate of basic microscopic motions $k_0$, by computing the
fraction of trajectories $\langle h_{\rm fold}\rangle_{\tau}\approx 1-\exp(-k^{(\alpha)}\tau)$
that fold within a fixed amount of time $\tau$ (with $k^{(\alpha)} \ll \tau^{-1} \ll k_0$). This
strategy offers two distinct advantages: (1) the evolutionary temperature $T_{\rm ev}$, which
governs the stringency of selection for efficient folding, can be controlled systematically; and
(2) estimates of folding efficiency via $\langle h_{\rm fold}\rangle_{\tau}$ can converge much
more rapidly than mean first passage time calculations employed in \citet{Mirny1998}.



 Our evolutionary simulations, conducted at moderate ``temperature'' $T_{\rm ev}=0.05\,/k_B$,
demonstrate that in fact many folding pathways can provide efficient access to a single native
state. It is therefore not at all self-evident that a particular, well-designed amino acid
sequence should arrive at its native structure via similar routes in full and \go-like versions of
the lattice heteropolymer model.


Using this method, we have generated hundreds of thousands of sequences which fold to given
structures (for example that of Fig. 1a) through a variety of folding mechanisms. This is the
ensemble of sequences we use in this paper. Further details of the evolutionary dynamics used to
generate these large ensembles of sequences will be given in a forthcoming publication \cite{Brian2}.

\bibliography{gonongo2}
\end{document}